\documentclass[aps,pra,twocolumn,groupedaddress]{revtex4}

\usepackage{graphicx}
\usepackage{epsfig}

\begin{document}

\DeclareGraphicsExtensions{.eps,.EPS,.jpg,.bmp}


\title{Raman laser spectroscopy of Wannier Stark states}


\author{G. Tackmann, B. Pelle, A. Hilico, Q. Beaufils, F. Pereira dos Santos }

\email[]{franck.pereira@obspm.fr}
\affiliation{LNE-SYRTE, UMR 8630 CNRS, Observatoire de Paris,
UPMC, 61 avenue de l'Observatoire, 75014 Paris, FRANCE}


\date{\today}

\begin{abstract}
Raman lasers are used as a spectroscopic probe of the state of atoms confined in a shallow 1D vertical lattice. For long enough laser pulses, resolved transitions in the bottom band of the lattice between Wannier Stark states corresponding to neighboring wells are observed. Couplings between such states are measured as a function of the lattice laser intensity and compared to theoretical predictions, from which the lattice depth can be extracted. Limits to the linewidth of these transitions are investigated. Transitions to higher bands can also be induced, as well as between transverse states for tilted Raman beams. All these features allow for a precise characterization of the trapping potential and for an efficient control of the atomic external degrees of freedom.

\end{abstract}


\maketitle



Cold atoms trapped in optical lattices have proven to be well suited for simulating solid-state systems, making possible the observation of Bloch oscillations \cite{salomon,nagerl1}, resonant tunneling \cite{arimondo,tinoresonant,nagerl} or the Mott insulator regime \cite{blochmott}. Besides, the precise knowledge and control of the atomic external degrees of freedom in these systems make them promising for applications such as metrology \cite{latticeclock,hsurm} or the development of inertial sensors \cite{cladeeuro,tinogravi}. In a recent article \cite{Beaufils} we showed that Raman pulses can be used to induce tunneling between neighboring sites of a 1D vertical lattice. The present article aims at providing a more detailed description of the system. In particular we use Raman spectroscopy to probe the energy structure of atoms trapped in a 1D vertical lattice. This study is motivated by recent proposals to use such a system for short range forces measurements \cite{proposal1,proposaltino,proposalinguscio,derevianko}, or for the realization of a compact gravimeter \cite{proposalisense,tinogravi,kovachy}. It is also of interest for any experiment using a shallow optical lattice, as it provides a comprehensive characterization of the system.

We consider atoms trapped in a vertical standing wave, created by a laser far detuned from resonance. This results in a periodic potential, which is superimposed to the gravitational potential in the vertical direction. The internal atomic structure is approximated by a two-level system with long lived states $|g\rangle$ and $|e\rangle$ with energy difference $h\nu_{eg}$. The total Hamiltonian of this system is given by
\begin{equation}\label{eq1}
\hat H=\hat H_{int}+\hat H_{l}+\hat H_{g},
\end{equation}
where $\hat H_{int}=h\nu_{eg}|e\rangle\langle e|$ represents the internal energy, $\hat H_{l}=U_0(1-\cos(2k_l\hat z))/2$ is the periodic lattice potential with lattice depth $U_0$, lattice wavenumber $k_l$ and vertical spatial coordinate $\hat z$, and $\hat H_g=m_ag\hat z$ represents the gravitational potential, where $m_a$ is the mass of the atom and $g$ is the gravity acceleration.

\begin{figure}[h]
      \begin{center}
     \includegraphics[width=8 cm]{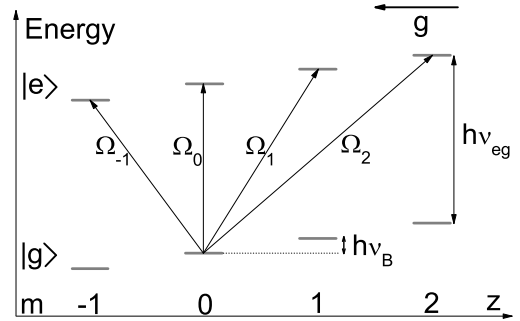}
\end{center}
    \caption{Wannier-Stark ladder of states and couplings between states by the probe laser}
    \label{wsladder}
\end{figure}

As known from solid state physics, the eigenstates of the external part
$\hat H_{l}+\hat H_{g}$, which is the sum of a periodic and a linear
potential, are given by Wannier-Stark (WS) states
\cite{wannierstark,WSobservation}. They form a so-called WS-ladder of
states $|W_{b,m}\rangle$, where $b$ is the discrete well index of the
Bloch band, which structures the eigenstates in the periodic lattice,
and the discrete quantum number $m$ is the well index, characterizing
the vertical position of the wave function $\langle
z|W_{b,m}\rangle$ and which labels the well containing its main peak in the limit of deep lattices ($U_0\gg E_r$). The energy
difference between adjacent lattice sites in the same band is simply the
potential energy difference between two neighboring wells
$h\nu_B=m_ag\lambda_l/2$, where $\lambda_l=2\pi /k_l$ is the lattice
wavelength, and $\nu_B$ is the Bloch frequency. Considering only the
bottom band ($b=0$) and adding the internal Hamiltonian leads to a new
WS-ladder like structure consisting of states $|g,W_m\rangle$ and $|e,W_{m'}\rangle$
(see figure \ref{wsladder}).


\begin{figure}[h]
      \begin{center}
     \includegraphics[width=8 cm]{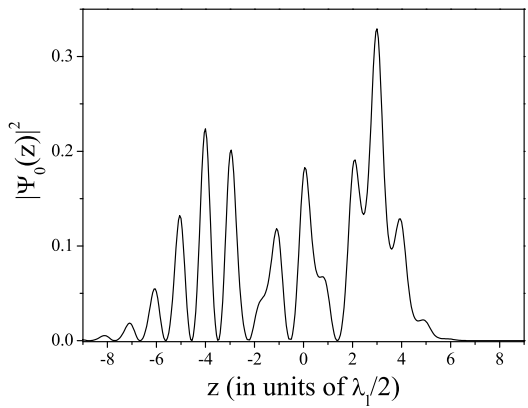}
\end{center}
    \caption{Spatial density distribution of the $|W_0\rangle$ wavefunction for a lattice depth of $U_0=1.6\,E_r$.}
    \label{fig:wsstate}
\end{figure}

In this structure, the application of a laser field resonant to $\Delta E/h=\nu_{eg}+\Delta m\nu_B$ with $\Delta m=0, \pm1, \pm2, \dots$ allows for coupling one state of this ladder to neighboring WS states with opposite internal state, and thereby for the determination of the clock frequency $\nu_{eg}$ and the local gravity $g$ in a spectroscopic measurement. In this, the inherent state labeling \cite{Borde} gives us a tool for the precise measurement of the external state by internal state detection. Coupling of the ladder states becomes apparent when adding a coupling Hamiltonian $\hat H_s=\hbar\Omega_{U_0=0}\cos(\omega t-k_s\hat z)|e\rangle\langle g|+H.c.$ to $\hat H$, where $\Omega_{U_0=0}$ is the Rabi frequency in absence of the lattice potential and $k_s=2\pi /\lambda_s$ the coupling laser's wave number~\cite{lemondewolf}. From this, the coupling strength for transitions between pairs of these states either in the same well ($\Delta m=m-m'=0$), or in neighboring wells ($\Delta m\neq 0$) is calculated to be \cite{lemondewolf}:
\begin{equation}\label{eq2}
\Omega_{\Delta m}=\Omega_{U_0=0}\langle W_m|\textnormal{e}^{ik_s\hat z}|W_{m'}\rangle .
\end{equation}
As we will see in more detail later in this paper, the lattice depth $U_0$ plays an important role for driving these transitions. In too shallow lattices, the atomic localization is too weak and Landau-Zener (LZ) tunneling occurs, which limits the WS state lifetime. For too deep lattices, the WS states are localized in only one well, which strongly limits the intersite coupling strength and thus compromises spectroscopy measurements. Figure \ref{fig:wsstate} illustrates the delocalization of the WS wavefunction at the depth of $1.6\,E_r$ we use in our experiment: it displays the spatial density probability of WS wavefunction $|\Psi_0(z)|^2=|\langle z|W_0\rangle|^2$, which extends over about 15 wells.

Intersite transitions can be realized choosing $\lambda_s$ in the optical domain, with a single laser connecting two different electronic states, such as the ground state and a metastable state of the optical clock transition studied in in \cite{bize}. Alternatively, they can be realized with two photon transitions between the two hyperfine ground states of alkali atoms. In this paper, we will focus on this latter case. For $^{87}$Rb, transitions between the ground and excited hyperfine levels $\left|g\right\rangle
=\left|5^{2}S_{1/2},F=1,m_{F}=0\right\rangle$ and
$\left|e\right\rangle
=\left|5^{2}S_{1/2},F=2,m_{F}=0\right\rangle$ can be driven using counterpropagating vertical Raman beams providing a frequency difference of $\nu_R=\nu_2-\nu_1$, that can be tuned around $\nu_{eg}=6.834\,$GHz. This transition implies a momentum transfer $k=k_{1}+k_{2}\approx 4\pi/\lambda_s$ with $\lambda_s=780$~nm replacing $k_s$ in equation~\ref{eq2}. Here, $\nu_{1,2}$ and $k_{1,2}$ are the respective frequencies and wavenumbers of the two Raman lasers.

The intersite coupling on this transition is discussed for different values of $\lambda_l$ in section~\ref{section1}. The experimental apparatus is then presented and the observed coupling strengths are compared to theoretical values. The achieved linewidth surpassing the Fourier limit by less than a factor two in the range up to 1.4~s of spectroscopic interrogation time and its limitations are presented and discussed in section~\ref{section2}. Finally, we show the observed longitudinal and transverse structures observed in our composed trap in section~\ref{section3}.

\section{Couplings}\label{section1}

\begin{figure}[h]
      \begin{center}
      \includegraphics[scale=0.7]{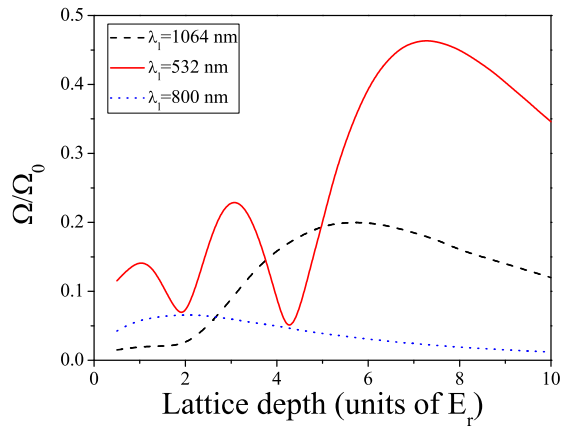}
\end{center}
    \caption{Coupling strengths $\Delta m=\pm1$ versus lattice depth for three different lattices wavelengths $\lambda_l=532$~nm, $\lambda_l=1064$~nm and $\lambda_l=800$~nm. $\lambda_s$=780~nm.}
    \label{fig:couplings1}
\end{figure}

The possibility to drive resolved intersite transitions strongly depends on the lattice wavelength and depth, as illustrated in figure \ref{fig:couplings1}. We calculated the coupling strengths as a function of the lattice depth (using a numerical calculation of the second term in equation 2, see also \cite{lemondewolf}) for $\Delta m=m'-m=\pm1$ transitions, for three different lattice wavelengths: close to resonance ($\lambda_l=800$~nm), far blue-detuned ($\lambda_l=532$~nm) and far red-detuned ($\lambda_l=1064$~nm). The choice of these wavelengths is motivated by the availability of powerful enough lasers ($\geq 20$~W in the far detuned cases, up to several watts close to resonance), allowing to reach a sufficient depth of a few recoil energies with a relatively large waist ($\approx$~1~mm). Such a waist is required for minimizing parasitic forces due to the dipolar potential gradient along the longitudinal direction when using such a system for high precision measurements (gravimetry, short range forces, fine structure constant \cite{biraben}...).

\begin{figure}[h]
      \begin{center}
      \includegraphics[scale=0.7]{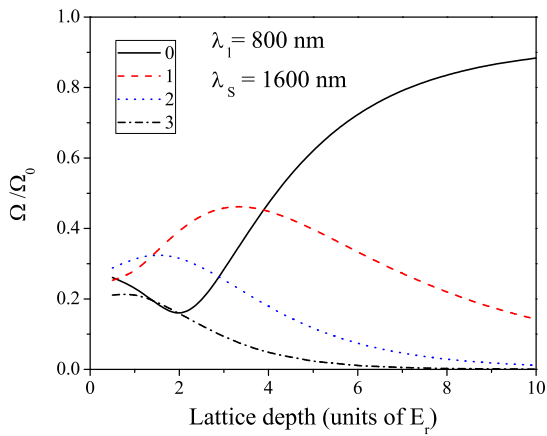}
\end{center}
    \caption{Coupling strengths $\Delta m=0,\pm1,\pm2,\pm3$ versus lattice depth for a lattice wavelength $\lambda_l=800$~nm and tilted Raman beams giving an effective wavelength of $\lambda_s=1600$~nm.}
    \label{fig:couplings2a}
\end{figure}

At sufficiently low depths, of a few $E_r$, we find coupling strengths of the same order of magnitude for the far blue and red detuned cases whereas the intersite coupling remains small for the close to resonance case. We find relatively large variations and modulations for the far detuned cases, with respect to the close to resonance case. Calculations performed for $\Delta m\geq2$ transitions showed similar behaviors.
Nevertheless, tilting by the same angle both Raman lasers with respect to verticality allows to preserve the direction of $k$ while reducing its magnitude, thus changing the coupling strengths. As an example, we plot in figure \ref{fig:couplings2a} the couplings in the close to resonance case for $k$ reduced by a factor of 2, which corresponds to an angle of $60^{\circ}$ with respect to verticality. Although we find larger coupling strengths, which oscillate versus lattice depths, couplings comparable to the far detuned cases for $\Delta m \geq 1$ are only reached at twice lower lattice depth.

\begin{figure}[h]
      \begin{center}
      \includegraphics[scale=0.7]{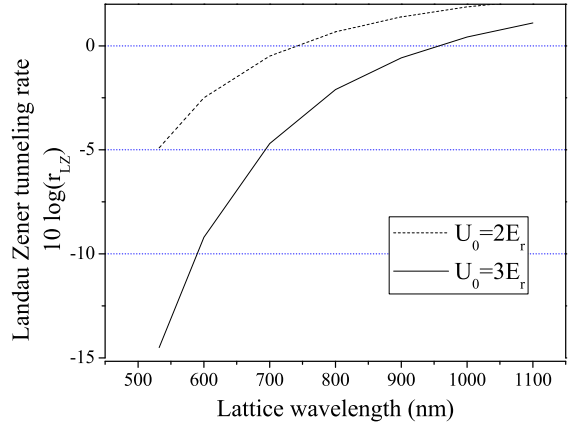}
\end{center}
    \caption{Landau Zener tunneling rate as a function of the lattice wavelength for $U_0=2$ and $3\,E_r$.}
    \label{fig:couplings2b}
\end{figure}

Another important parameter depending on the lattice wavelength is the tunneling rate out of the bottom band, which limits the lifetime of the atoms in shallow lattices. For an estimate of this rate, we use the Landau-Zener formula. Figure \ref{fig:couplings2b} displays the calculated rates as a function of the lattice depth (in units of recoil energies) for low depths of 2 and 3~$E_r$. The LZ tunneling rate remains small in the far blue detuned case, even for lattices as shallow as $2\,E_r$, whereas it becomes comparable to $1$ s$^{-1}$ for a lattice depth $U_0$ between 2 and $3\,E_r$ in the close to resonance case. Operation at $\lambda_l=532$~nm thus appears more appealing as one can combine large lifetimes and good couplings for large site offsets $\Delta m$.
\newline

\begin{figure}
    \centering
        \includegraphics[scale=0.45]{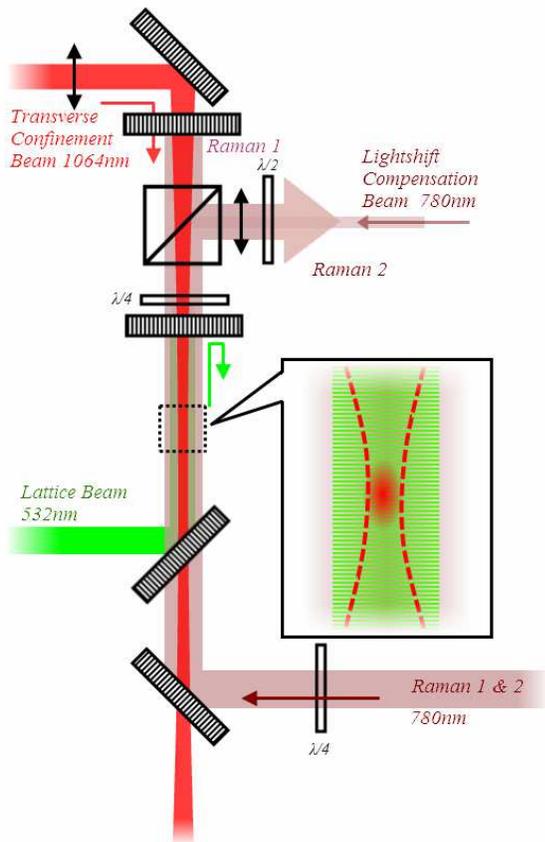}
    \caption{Experimental setup. The laser beams for optical trapping (lattice at 532 nm and transverse confinement at 1064 nm) and Raman spectroscopy (780 nm) are superimposed using dichroic optics.}
        \label{setup}
\end{figure}

We have compared these calculations with measurements corresponding to the blue detuned case. Our system \cite{Beaufils} consists in laser-cooled $^{87}$Rb atoms loaded in the first band of a vertical one-dimensional optical lattice, created by a single mode frequency doubled Nd:YVO$_{4}$ laser ($\lambda_{l}=532$~nm, maximal power 12~W) with a waist of about 700~$\mu$m. As this blue detuned standing wave does not provide transverse confinement, a red detuned ($\lambda = 1064$~nm, beam waist $200\,\mu$m) Yb fiber laser is superimposed to the lattice (see figure \ref{setup}). The difference in the waists of the two lasers allows to reduce inhomogeneities in the lattice depth due to the transverse extension of the atomic sample.
Before being transferred into this mixed dipole trap, about $10^{7}$ atoms are loaded in a 3D-Magneto-Optical trap (MOT) and cooled down to $2\,\mu$K with a far detuned molasses. The dipole trap lasers are switched on either at the end of this cooling phase, or at the beginning of the MOT sequence. Switching off the molasses lasers leaves about 1 $\%$ of the atoms trapped in the mixed trap with a lifetime of about 1 s. These atoms initially distributed in all the Zeeman sublevels of $\left|5^{2}S_{1/2},F=2\right\rangle$ are then depumped to $\left|5^{2}S_{1/2},F=1\right\rangle$ before being optically pumped ($98 \%$ efficiency) on the $\left|5^{2}S_{1/2},F=1\right\rangle \rightarrow \left|5^{2}P_{3/2},F=0\right\rangle$ transition to the $\left|5^{2}S_{1/2},F=1,m_{F}=0\right\rangle$ Zeeman sublevel, which is sensitive to stray magnetic fields only to second order. After being released from the optical trap, atoms fall for about 140 ms before reaching the detection zone located at the bottom of the vacuum chamber. The detection scheme is based on a time of flight measurement similar to the ones used in atomic clocks and inertial sensors. It allows to measure by fluorescence the atomic populations in the two hyperfine states $F=1$ and $F=2$, denoted $N1$ and $N2$, respectively \cite{detection}, from which we derive the transition probability $P=N2/(N2+N1)$. The Raman transitions are driven by two counterpropagating, circularly polarized beams at $780$~nm, detuned from the atomic transition by about $-3$~GHz, and aligned along the direction of the optical trap beams. These are collimated with a $1/e^{2}$ radius of $1$ cm, ensuring a good intensity homogeneity along the transverse size of the trap.

In order to determine the coupling strengths, a Raman spectrum is first scanned by measuring the transition probability as a function of the frequency difference between the Raman lasers $\nu_R$. For such scans, the intensities in the Raman laser beams are $0.25$ and $0.54$ mW$/$cm$^{2}$, and the duration of the Raman pulse is 8 ms.
This ratio between the Raman intensities is chosen to cancel the differential light shift they induce on the frequency of the hyperfine transition \cite{Raman}.

\begin{figure}
    \centering
        \includegraphics[scale=0.55]{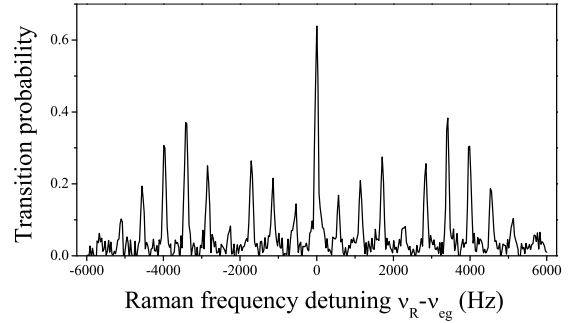}
    \caption{Raman spectrum showing the transition probability as a function of the Raman frequency from a lattice depth of $1.6\,E_r$. The resonances separated by the Bloch frequency $\nu_{B}\approx 569$~Hz are the signature of intersite transitions.}
        \label{spectrum}
\end{figure}

\begin{figure}
    \centering
        \includegraphics[scale=0.65]{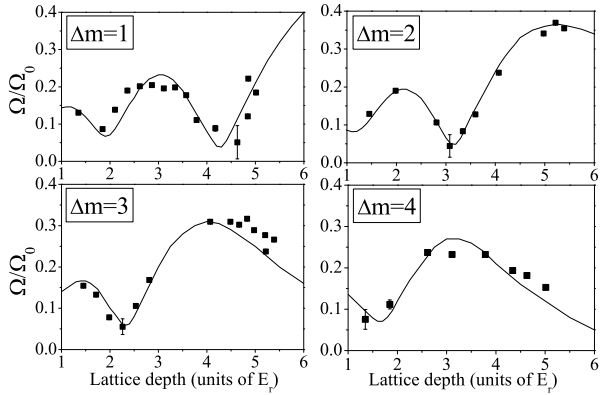}
    \caption{Normalized Rabi frequencies measured as a function of the lattice depth, for $\Delta m=1,2,3,4$. The normalization factor $\Omega_{U_0=0}$ is an adjustable parameter. Solid lines are the result of numerical evaluation of equation \ref{eq2}.}
        \label{fig:couplings}
\end{figure}

We observe multiple resonances, corresponding to transitions between the two hyperfine levels at Raman frequencies equal to the hyperfine splitting plus or minus an integer number $\Delta m$ of Bloch frequencies ($\nu_{B}\approx 569$~Hz in our system). The difference in the peaks heights is due to the difference in the coupling strengths. We then fix the Raman frequency difference at the center of each peak, record a Rabi oscillation pattern by measuring the transition probability as a function of the pulse length, from which we extract the Rabi frequency. We repeat this procedure for different lattice laser power values. The measured coupling strengths of the $\Delta m=0,1,2,3$ transitions are plotted in figure \ref{fig:couplings} as a function of the lattice laser power. The results have been normalized and the relation between lattice laser power and actual lattice depth has been adjusted for a better match with the theoretical predictions. Data points corresponding to minima in the couplings have larger error bars as inhomogeneities in the Rabi frequencies damp so heavily the Rabi oscillations that almost no oscillations are observed. We attribute this inhomogeneity to the transverse spread of the atoms, which experience different lattice depths due to the Gaussian profile of the lattice beam.

The good agreement between the measurements and the theoretical predictions allow us to determine the depth with a resolution on the order of $0.1\,E_r$ from the direct comparison of the relative amplitudes of the peaks. Alternative techniques for the determination of the lattice depth are not as accurate here: diffraction in the thick grating limit \cite{Ovchinnikov} creates sidebands in the velocity distribution which can hardly be resolved due to the width of the initial velocity distribution of the order of $2.5 v_r$, and parametric excitation gives rise to very wide resonances due to the complete anharmonicity of the lattice potential.

\section{Linewidth}\label{section2}

We then investigated the question of the linewidth of such transitions. Various effects are expected to contribute to the broadening of the transitions and to ultimately limit the minimally attainable linewidth. One of them is the differential light shift (DLS) induced by the trapping laser beams. This effect is dominated by the light field of the transverse dipole trap, as atoms are trapped at the intensity maxima of the 1064~nm beam and observe maximal DLS$_{IR}$ at its center. By performing microwave spectroscopy on the $|5^{2}S_{1/2},F=1,m_F=0\rangle \rightarrow |5^{2}S_{1/2},F=2,m_F=0\rangle$ transition, we measure a shift in the center of the line of about 3~Hz/W, and a broadening of about 2~Hz/W, which gives a limit to the linewidth of 3~Hz at the 1.5~W we typically use. We have also measured the shift of the line induced by the lattice laser beam and found a much smaller effect of 0.4~Hz at full power.

\begin{figure}[h]
    \centering
        \includegraphics[scale=0.7]{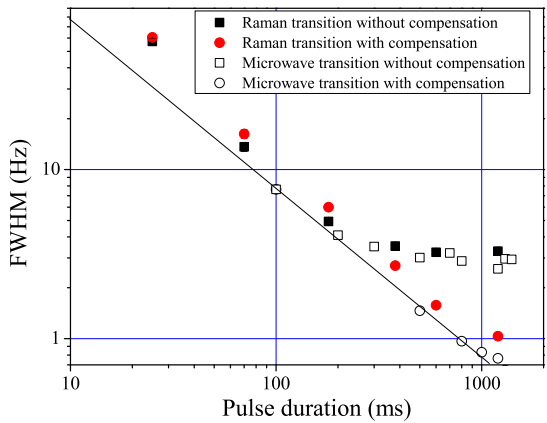}
    \caption{Linewidth (FWHM) of transitions between the hyperfine states driven by a microwave probe in the same lattice well or by counterpropagating Raman lasers with $ \Delta m = 3 $, as a function of the pulse duration $\tau$. For each different pulse duration, the probe's Rabi frequency $\Omega_3$ is adjusted so that $\Omega_3 \tau = \pi$. The solid line shows the theoretical Fourier-limited FWHM of a pulse of duration $\tau$.  }
        \label{fig:linewidth}
\end{figure}

\begin{figure}[h]
    \centering
        \includegraphics[scale=0.8]{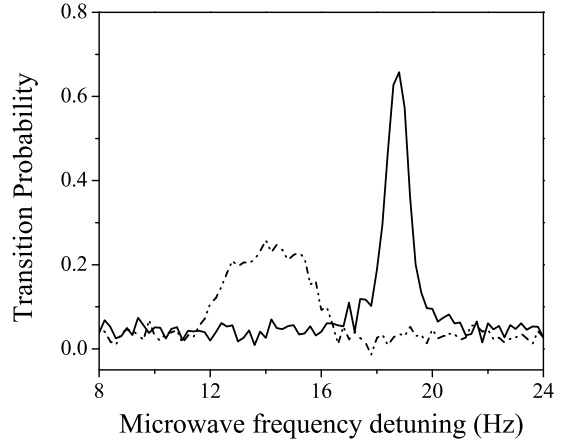}
    \caption{Transition probability as a function of the microwave detuning with (continuous line) or without (dotted line) the DLS compensation beam (see text), for a microwave pulse of $\tau = 1.4$~s. The compensated transition's detuning to the hyperfine frequency of $18.7$~Hz is the Zeeman quadratic shift due to a bias field of $180$~mG. The mean DLS imposed by the transverse trapping laser is $-4.6$~Hz. }
    \label{fig:ls}
\end{figure}

This broadening is illustrated in figure \ref{fig:linewidth}, which shows the evolution of the linewidth as a function of the duration of the microwave pulse, where the microwave power has been adjusted for optimal transfer at resonance (which corresponds to the case of a $\pi$ pulse), as well as in figure \ref{fig:ls}, which shows as a dotted line the microwave spectrum corresponding to a pulse duration of 1.4 s.

The DLS$_{IR}$ induced by the transverse trapping laser can be compensated thanks to an additional laser beam with a blue detuning for the $|e\rangle$ state and a red detuning for the $|g\rangle$ state \cite{kaplan}. For that purpose, a small fraction of one of the two Raman beams is used, with an additional detuning of 80~MHz in order to prevent undesired Raman transitions. This beam is superimposed with the transverse trapping laser beam and its size, position and power are adjusted to reduce the broadening of the microwave transition. For a transverse trapping laser power of 1.5~W, the differential light shift is compensated with a power of $12$~nW. Figure \ref{fig:ls} displays as a continuous line the microwave spectrum for optimal compensation.

The linewidths of the Raman transitions are displayed in figure \ref{fig:linewidth}, for uncompensated (resp. compensated) DLS$_{IR}$, as squares (resp. circles). The ratio between the intensities of the two Raman lasers is set to cancel (on average) the net differential light shift they induce. However, due to differences in the spatial modes of the two Raman lasers and parasitic reflections, this compensation is not perfect, which leads to a broadening of the hyperfine transition. As this broadening is proportional to the total Raman intensity, the linewidth is proportional to the Rabi frequency of the transition, only increased with respect to the Fourier-limited microwave transition by a constant factor of about 1.5. Atom loss prevents us from driving longer transitions. Nevertheless, our system allows us to achieve a spectroscopic resolution of about 1~Hz, which can be of interest for selecting atoms in a single site of the lattice, as demonstrated in \cite{siteselect} with less resolved transitions.

In order to study the short term sensitivity of our system, we performed a spectroscopic measurement of the Bloch frequency by measuring alternately the frequency of the $\Delta m=+3$ and $\Delta m=-3$ and calculating the difference to cancel any shift of the hyperfine clock frequency. We obtained a statistical uncertainty on the Bloch frequency of $2 \times 10^{-5}$ in relative value after 1~s of integration, which is a factor of 3 better than the previously reported sensitivity in \cite{Beaufils} using Ramsey spectroscopy. The best sensitivity reported for a trapped accelerometer was $1.4 \times 10^{-7}$ in relative value after one hour measurement time \cite{tinogravi}, which corresponds to an equivalent relative short term sensitivity of $9 \times 10^{-6}$ at 1 s.

\section{Probing the lattice's longitudinal and transverse structures}\label{section3}

\begin{figure}[h]
    \centering
        \includegraphics[scale=0.8]{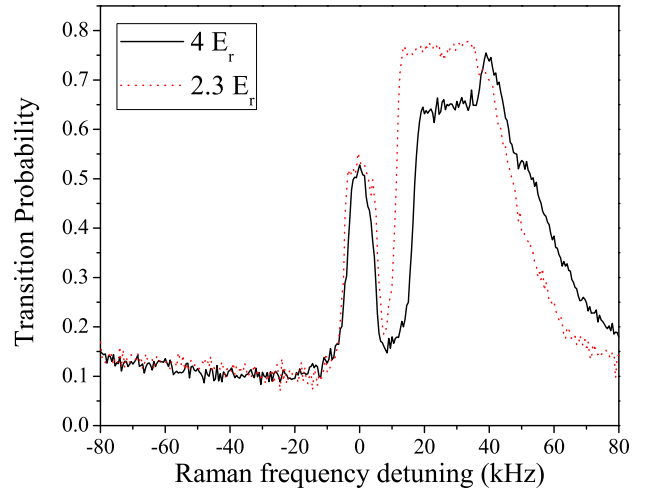}
    \caption{Transition probability as a function of Raman frequency. The broad peak at vanishing relative Raman laser detuning ($\nu_R\nu_{eg}=0$ corresponds to unresolved intersite transitions in the same lattice band. The large structure arising between 10 and 50~kHz is due to a coupling from the bottom to the first excited band. }
        \label{fig:scan}
\end{figure}

Performing the Raman scan over a larger frequency range reveals additional features. Figure \ref{fig:scan} displays such a spectrum for two different lattice depths of 4 and $2.3\,E_r$ (measured by Raman spectroscopy as described in the last paragraph of section \ref{section1}). Note the large blue sideband, which corresponds to transitions from the bottom band to the first excited band. The absence of red sideband indicates that the trapped atoms have been loaded in the bottom band and that upper bands are not populated.

\begin{figure}[h]
    \centering
        \includegraphics[scale=0.8]{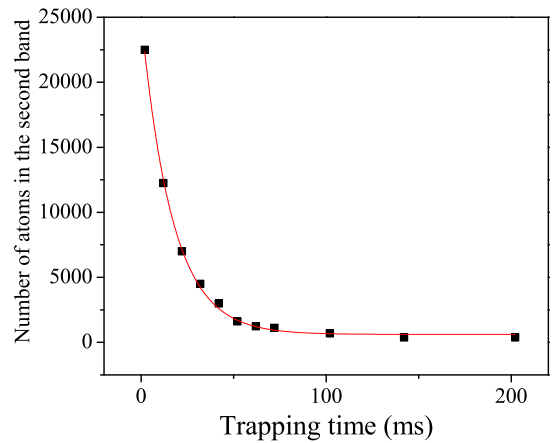}
    \caption{Population in the first excited band as a function of trapping time. The solid line is an exponential decay fit to the data from which we extract a lifetime of 16~ms.}
        \label{fig:lifetime}
\end{figure}

We have measured the lifetime of the atoms in the first excited band for a lattice depth of $4\,E_r$. A Raman pulse of 2~ms detuned by 30~kHz transfers 60~$\%$ of the atoms initially in $F=1$ in the excited band in the $F=2$ state. When increasing the delay between the Raman pulse and the turning off of the lattice laser, and measuring the number of the atoms that have remained trapped, we observe a decay in the number of atoms in $F=2$. Corresponding data are displayed in figure \ref{fig:lifetime}, from which we extract an exponential lifetime of 16~ms. This relatively short lifetime explains that for trapping times as large as several hundreds of ms, only the bottom band is populated. Loading the shallow lattice from the initial thermal distribution simply selects atoms loaded in the bottom band.

\begin{figure}[h]
    \centering
        \includegraphics[scale=0.8]{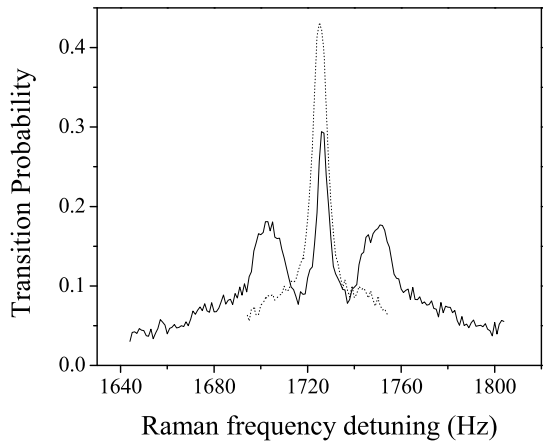}
    \caption{Dotted line: Transition probability as a function of Raman frequency around the $\Delta m=+3$ transition, when the Raman lasers wavevector is slightly misaligned from the lattice's one. The two sidepeaks correspond to intersite transitions involving a change in the transverse vibrational state. Solid line: the Raman lasers wavevector is aligned with the lattice's one.  }
        \label{fig:transverse}
\end{figure}

Transitions between vibrational states along the transverse directions can also be induced by the Raman lasers, provided that their effective wavevector projection along the transverse direction is not null. Such transitions are exploited for instance for Raman sideband cooling \cite{sidebandcooling1,sidebandcooling2}. To do so, we simply tilt the Raman retroreflecting mirror. Figure \ref{fig:transverse} displays zooms of the Raman spectrum close to the $\Delta m=+3$ transition, for a Raman pulse of 400~ms, for $\vec k$ perfectly vertical (continuous lines) and tilted by a few mrad (dotted line). We find red and blue sidebands about 25~Hz apart from the carrier, which correspond to transitions $\Delta n=\pm1$, where $n$ is the index of the transverse vibrational level. We find equal amplitude for both sidebands, indicating that atoms are distributed among many such $n$ states. The transverse temperature was independently measured by time of flight fluorescence imaging to be $1\,\mu$K. In addition, the sidebands are significantly broadened with respect to the carrier. This broadening is attributed to the anharmonicity of the potential, the depth of the transverse dipole trap being only about 4 times the average transverse kinetic energy.

\section{Conclusion}

We have investigated the possibility to drive intersite transitions in an optical lattice using Raman transitions. We have shown that good couplings between neighboring wells and high resolution Raman spectroscopy could be achieved in a composite trap, formed by a shallow blue detuned vertical lattice combined with a 1064 nm laser progressive wave for transverse confinement. Broadening due to the inhomogeneity of the differential light shift of the trapping laser can be prevented using an additional laser beam for differential light shift compensation. Raman transitions allow for a precise determination of the parameters of the shallow 1D lattice (depth, band filling, radial oscillation frequency ...). This spectroscopic tool allows for a sensitive determination of the Bloch frequency, for instance using the Ramsey type interferometer scheme demonstrated in \cite{Beaufils}, and can be used for the measurement of short range forces, when such an interferometer is created close to a surface \cite{proposal1}.
\newline

\begin{acknowledgments}

This research is carried on within the project iSense, which acknowledges the financial support of the Future and Emerging Technologies (FET) programme within the Seventh Framework Programme for Research of the European Commission, under FET-Open grant number: 250072. We also gratefully acknowledge support by Ville de Paris (Emergence(s) program) and IFRAF. G.T. thanks the Intercan network and the UFA-DFH for financial support.

Helpful discussions with P. Wolf, S. Pelisson, M-C. Angonin and R. Messina are greatfully acknowleged.

\end{acknowledgments}


\begin{thebibliography}{0}
\expandafter\ifx\csname natexlab\endcsname\relax\def\natexlab#1{#1}\fi
\expandafter\ifx\csname bibnamefont\endcsname\relax
  \def\bibnamefont#1{#1}\fi
\expandafter\ifx\csname bibfnamefont\endcsname\relax
  \def\bibfnamefont#1{#1}\fi
\expandafter\ifx\csname citenamefont\endcsname\relax
  \def\citenamefont#1{#1}\fi
\expandafter\ifx\csname url\endcsname\relax
  \def\url#1{\texttt{#1}}\fi
\expandafter\ifx\csname urlprefix\endcsname\relax\def\urlprefix{URL }\fi
\providecommand{\bibinfo}[2]{#2}
\providecommand{\eprint}[2][]{\url{#2}}

\end{thebibliography}


\begin{thebibliography}{30}




\bibitem{salomon} M. Ben Dahan, E. Peik, J. Reichel, Y. Castin, and C. Salomon,
Phys. Rev. Lett. \textbf{76}, 4508 (1996)

\bibitem{nagerl1} M. Gustavsson, E. Haller, M. J. Mark, J. G. Danzl, G. Rojas-Kopeinig, and H.-C. N\"agerl, Phys. Rev. Lett. \textbf{100}, 080404 (2008)

\bibitem{arimondo} C. Sias, A. Zenesini, H. Lignier, S. Wimberger, D. Ciampini, O. Morsch, and E. Arimondo,
Phys. Rev. Lett. \textbf{98}, 120403 (2007)

\bibitem{tinoresonant} V. V. Ivanov, A. Alberti, M. Schioppo, G. Ferrari, M. Artoni, M. L. Chiofalo, and G. M. Tino, Phys. Rev. Lett. \textbf{100}, 043602 (2008)

\bibitem{nagerl} E. Haller, R. Hart, M. J. Mark, J. G. Danzl, L. Reichs\"ollner, and H.-C. N\"agerl, Phys. Rev. Lett. \textbf{104}, 200403 (2010)

\bibitem{blochmott} M. Greiner, O. Mandel, T. Esslinger, T.W. H\"ansch, and
I. Bloch, Nature (London) \textbf{415}, 39 (2002)


\bibitem{latticeclock} M. Takamoto, F.-L. Hong, R. Higashi, and H. Katori,
Nature (London) \textbf{435}, 321 (2005)

\bibitem{hsurm} P. Clad\'{e}, E. de Mirandes, M. Cadoret, S. Guellati-Kh\'{e}lifa, C. Schwob, F. Nez, L. Julien, and F. Biraben, Phys. Rev. Lett. \textbf{96}, 033001 (2006)

\bibitem{cladeeuro} P. Clad\'{e}, S. Guellati-Kh\'{e}lifa, C. Schwob, F. Nez, L. Julien, and F. Biraben, Europhys. Lett. \textbf{71} (2005), 730

\bibitem{tinogravi} N. Poli, F.-Y. Wang, M. G. Tarallo, A. Alberti, M. Prevedelli, and G. M. Tino, Phys. Rev. Lett. \textbf{106}, 038501 (2011)

\bibitem{Beaufils} Q. Beaufils, G. Tackmann, X. Wang, B. Pelle, S. P\'{e}lisson, P. Wolf, and F. Pereira dos Santos, Phys. Rev. Lett. \textbf{106}, 213002 (2011)

\bibitem{proposal1} P. Wolf, P. Lemonde, A. Lambrecht, S. Bize, A. Landragin, and A. Clairon, Phys. Rev. A \textbf{75}, 063608 (2007)

\bibitem{proposaltino} G. Ferrari, N. Poli, F. Sorrentino, and G. M. Tino, Phys. Rev. Lett. \textbf{97}, 060402 (2006)

\bibitem{proposalinguscio} I. Carusotto, L. Pitaevskii, S. Stringari, G. Modugno, and M. Inguscio, Phys. Rev. Lett. \textbf{95}, 093202 (2005)

\bibitem{derevianko} A. Derevianko, B. Obreshkov, and V. A. Dzuba, Phys. Rev. Lett. \textbf{103}, 133201 (2009)

\bibitem{proposalisense} M. de Angelis \textit{et al.} , Proceedings of FET 11 conference to be published in Physics Procedia

\bibitem{kovachy} T. Kovachy, J. M. Hogan, D. M. S. Johnson, and M. A. Kasevich, Phys. Rev. A \textbf{82}, 013638 (2010)

\bibitem{wannierstark} G. Nenciu, Rev. Mod. Phys. \textbf{63}, 91 (1991)

\bibitem{WSobservation} S. R. Wilkinson, C. F. Bharucha, K. W. Madison, Qian Niu, and M. G. Raizen, Phys. Rev. Lett. \textbf{76}, 45124515 (1996)


\bibitem{Borde} Ch. J. Bord\'{e},  Physics Letters, A140, 10-12 (1989)

\bibitem{lemondewolf} P. Lemonde and P. Wolf, Phys. Rev. A \textbf{72}, 033409 (2005)

\bibitem{bize} L. Yi, S. Mejri, J. J. McFerran, Y. Le Coq, and S. Bize, Phys. Rev. Lett. \textbf{106}, 073005 (2011)


\bibitem{biraben} R. Bouchendira, P. Clad\'{e}, S. Guellati-Kh\'{e}lifa, F. Nez, and F. Biraben, Phys. Rev. Lett. \textbf{106}, 080801 (2011)












\bibitem{detection} J. Le Gou\"et, T. E. Mehlst\"aubler, J. Kim, S. Merlet, A. Clairon,
A. Landragin, and F. Pereira Dos Santos, Appl. Phys. B \textbf{92}, 133
(2008)


\bibitem{Raman} D.S. Weiss, B.C. Young and S. Chu, Appl. Phys. B \textbf{59}, 217 (1994)
















\bibitem{Ovchinnikov} Yu. B. Ovchinnikov, J. H. M\"uller, M. R. Doery, E. J. D. Vredenbregt, K. Helmerson, S. L. Rolston, and W. D. Phillips, Phys. Rev. Lett. \textbf{83} 284 (1999)

\bibitem{kaplan} A. Kaplan, M. F. Andersen, and N. Davidson, Phys. Rev. A \textbf{66}, 045401 (2002)

\bibitem{siteselect} M. Karski, L. Forster, J.M. Choi, A. Steffen,
N. Belmechri, W. Alt, D. Meschede, and A. Widera , New J. Phys. \textbf{12} 065027 (2010)

\bibitem{sidebandcooling1} S. E. Hamann, D. L. Haycock, G. Klose, P. H. Pax, I. H. Deutsch, and P. S. Jessen , Phys. Rev. Lett. \textbf{80} 4149 (1998)

\bibitem{sidebandcooling2} H. Perrin, A. Kuhn, I. Bouchoule, and C. Salomon, Europhys.
Lett. \textbf{42}, 395 (1998)

\end{thebibliography}

\end{document}